July 9, 2015

# Schrödinger-Milne Big Bang—Creating a 'Universe of Threeness'

Geoffrey F. Chew

*Theoretical Physics Group*
*Physics Division*
*Lawrence Berkeley National Laboratory*
*Berkeley, California 94720, U.S.A.*

**Summary**

A Schrödinger-evolving forward-lightcone-interior 'Milne' universe ('SMU') is governed by 'centered-Lorentz' (CL) symmetry—that of a 9-parameter Lie group with a 6-parameter SL(2,c) 'exterior' and a 3-parameter 'quality-space' *center*. 'Reality' resides in current densities of electric charge and energy-momentum--the Dalembertian of an SMU-*ray-specified* classical *retarded* Lorentz-tensor field with $2^2$ *electromagnetic* and $3^2$ *gravitational* components.

Nine conserved Dirac momenta comprise the CL algebra. We here propose a Dirac self-adjoint CL-invariant *Hamiltonian*—kinetic energy plus electromagnetic and gravitational potential energy—to evolve the SMU *ray* from a featureless beginning. Illuminated here via *discreteness* of electric charge are baryon number ('nuclear forces' arising from 'almost-screened' electric charge), Bohm's 'hidden variables'--dark matter and dark energy--*and* three generations of 'elementary' fermions.

**Introduction**

Dirac's nonrelativistic quantum theory [1] was based on Hilbert-space self-adjoint operators—Dirac *coordinates* and Dirac *momenta* plus a *Hamiltonian* which commuted with the Dirac-*momentum*-represented *algebra* of a Euclidean *Lie symmetry group*. Dirac *coordinates* represent the symmetry-group's *manifold*. Ray evolution was according to a Schrödinger first-order differential equation. Such quantum theory has for 'relativistic' physics been obstructed by *absence* of *unitary finite*-dimensional Lorentz-group representations.

The Gelfand-Naimark (GN) *unitary* Hilbert-space *infinite*-dimensional representations, [2] although inapplicable to a physics focused on 'objective reality'—i.e., on *particles*—apply *quantum-cosmologically* to a Schrödinger-Milne Planck-scale-originated--'big-bang'—a continuously-evolving universe (SMU) whose galaxies presently contain more 'dark matter' than particulate. *Non*-galactic *non*-particulate *unphysical* energy continues at present to exceed that of galaxies.

Milne's universe [3] situates *inside* a Lorentz-Minkowski forward lightcone—a 4-dimensional manifold whose boundary locates *outside* Milne's universe-occupied *submanifold*. A positive Lorentz-invariant 'age', $\tau \geq \tau_0 > 0$, of any location within Milne's universe is *defined* to be its 'Minkowski distance' (*not* a 'Riemann geometrical distance') from the lightcone vertex. Universe *age* at *big bang* was $\tau_0$.

Milne's geometrical 3-space at *any* fixed age is hyperbolic—negatively curved in Riemann sense as suggested by the surprising astronomically-observed Nobel-acknowledged correlation between luminosity distance and redshift. [4] The present paper will 'embellish' GN's unitary Hilbert-space Lorentz-group representation with discrete electric charge. A 9-parameter SMU Lie symmetry group, here denoted 'CL', then enriches Milne's 6-dimensional SL(2,c) 'universe-exterior' by a 3-dimensional group *center* that, in spanning the universe's 'quality space', recognizes discreteness *both* of electric charge *and* of energy.

Through an approximate relation between redshift and the (geometrical) distance light travels through Milne's negatively-curved noncompact 3-space, *age* is astronomically estimate-able. *Present* age is approximated by the reciprocal of astronomy's 'Hubble constant'. [4]

'Cosmological photons' ($\gamma_c$), first created at extremely early SMU ages, enjoy a 'semi-foundational' status in SMU's unification of electromagnetism and gravity--a GN-dependent unification *without* 'gravitons' that attributes the strength and 'short range' of (present-age) 'nuclear forces' to any nucleon (of unit baryon number) being a *composite* of nine electrically-charged 'quantum-universe constituents'--*quc*s. *One* of many important outcomes of *discrete-charge-screening* is 'short-range nuclear forces'. A principal task of the present paper is to define '*quc*'. An electron comprises three electrically-charged *quc*s, and a photon or neutrino two. Dark-matter-composing *quc*s are *chargeless*.

Decades of thought about alternative possibilities, plus attention to Occam's principle, have led the author here to propose SMU birth *without any* 'particles' at a *Planck scale* age. *But* there *was*, at SMU's start, a *huge* (although finite) energy which has diminished subsequently by *inverse age*-proportionality of each *quc*'s energy. The here-proposed set of SMU-composing *quc*s is finite and *fixed*, although ginormous.

The SMU ray at age $\tau \geq \tau_0 > 0$—a sum of ('tensor') products of single-*quc* Hilbert vectors--is an *indefinitely-differentiable* function of $\tau$—without singularities at any age greater than or equal to the positive starting-age $\tau_0$ whose value, together with the (permanently-fixed) *number* of (different) *quc*s and an electric-charge unit, provides SMU's foundation.

At *any* age (above $\tau_0$), ray expectations of certain here-defined self-adjoint Hilbert-space operators specify a 'reality' consistent *at present* and *recent* ages with Karl Popper's humanistic *classical* meaning for 'physical measurement'. But SMU's *age-varying* 3-space curvature *requires* that an inherently-approximate *Euclidean* meaning for 'quantum physics' be distinguished from the meaning of 'quantum-cosmology'.

Specified here is a (Dirac-sense) self-adjoint (even though 'cosmological') Hilbert-space Hamiltonian operator whose *potential energy* generates 'creative' ray dynamics (*crd*). Kinetic energy 'perpetuates' the creativity of evolution. Appendix A proposes a creation-bereft (particle-absent) *initial* ray *all* of whose *quc*s carried positive energy.

SMU's attribution, through *mathematical* language, of *cosmological* meaning to *Hamiltonian* kinetic and potential energy as well as to 'momentum', 'angular momentum', 'electric charge' and 'energy'—*all* notions uncovered by human physics--has amazed the author. Is it conceivable that the human-uncovered *mathematical* notions of 'Riemannian geometry', 'Lie-group algebra', 'Hilbert space', 'fiber bundle', 'differential equation' and 'Mersenne prime' play 'cosmological' roles that transcend humanity? This paper supposes such to be the case.

Milne's hyperbolic *three*-dimensional space at *fixed* universe age is CL-*invariantly* metricized and thereby, although (negatively) *curved*, endowed with an unambiguous (positive, Lorentz-frame-independent) *shortest* path (geodesic) between any pair of spatial locations. [4] Completely *age-determined* (without dependence on 3-space location), Milne's spatial curvature ignores energy *distribution*--curvature being initially at Planck scale and diminishing thereafter—paralleling diminishment with advancing age of *any* and *all quc* energies. Hubble's 'constant' approximates *both* the inverse of present-universe age *and* the current magnitude of (negative) 3-space curvature. [4]

Above emphasized is SMU governance by the symmetry of a 9-parameter CL 'centered-Lorentz' Lie group whose (below-compactified) *center* is 3-parameter $U(1) \times SL(2,c,D)$. [4] Here the symbol D denotes 2-parameter *left-acting diagonal* complex unimodular 2×2 matrices. The (non-compact) CL exterior is 6-parameter *right-acting* $SL(2,c)$. Formula (8) below, within our main text, shows how 'left-right' distinction is achieved by GN's remarkable Hilbert space (never associated by its discoverers to cosmology).

CL extends $SL(2,c)$ by a *trio* of single-parameter CL-commuting subgroups. *Algebra* extension from 6 to 9 elements defines, firstly, additively-conserved (*ac*) discrete and superselected *electric charge* with units proportional to $(\hbar c)^{\frac{1}{2}}$, secondly, a discrete 'semi-superselected' attribute dubbed 'chirality' with $\hbar/2$ units and, finally, discrete *ac* energy with $\hbar/2\tau$ units. As later elaborated, the universe's 3-dimensional 'quality space' is spanned by CL's center.

Energy discreteness does *not* mean Hamiltonian diagonalization. (Spacing between successive possible single-*quc* energies is $\hbar/2\tau$.) The quantum-physics notion of 'stationary state' enjoys *no* cosmological meaning. SMU displays 'onflow'—never-ending *continuous* development of 'newness'.

The 6-element ('acting from the right') exterior CL algebra defines angular momentum and momentum. Although the former (when Stone-Dirac represented) is discrete, the latter is not. *Non-compactness* of CL exterior associates in SMU to *continuous* spectra for a (3-vector) trio of Hilbert-space self-adjoint (Dirac) *quc-momentum* operators, whose non-Euclidean (cosmological) *failure to mutually commute* lacks physics precedent. Physics meaning for the term 'boost' is *absent* from SMU's cosmological vocabulary. An SMU Stone-Dirac *momentum* 3-vector operator generates displacements in Milne's curved 3-space.

Conservation of electric charge and chirality, as well as of angular momentum, is SMU (Noether) sustained by *unitary* Hilbert-space CL representation *and* CL-invariance of Hamiltonian. Despite *inverse* age-proportionality of *all quc* energies and momenta (*not* of 'particle masses', whose significance is *physical--not* cosmological), the Cl-central algebra adjoins a *positive ac energy-integer* to charge and chirality integers--thereby defining a (permanent) 'complete set' of *different* SMU constituents.

Wigner's *flat* 3-space (10-parameter) 'Poincaré group', foundation for quantum field theory (QFT) and the S matrix, *fails* to admit (Dirac-required) *unitary* Hilbert-space representation. Having been apprised by D. Finkelstein (private communication) that the Lorentz group 'Inonu-contracts' to the Euclidean group in a $\tau \to \infty$ limit, we have come to regard QFT as a '3-space-flattened' micro-macro-scale *approximation* which, for *human-history* values of $\tau/\tau_0$ (~$10^{60}$), is adequate *for human physics purposes* (FHPP, imitating John Bell's acronym) although *not* for all purposes of Schrödinger-Milne *cosmology*. (The author believes Bell, philosopher as well as physicist but *not* cosmologist, to have regarded humanity's 3-space as FAPP Euclidean.)

Physics, able to 'Popper-define' particles but *not quc*s, is *unable* to describe 'dark matter'. Bewilderingly (to the author), Maxwell *classical*-electromagnetic theoretical physics, through 4-vector electric-charge current density with *discrete* electric charge, plus symmetric-tensor energy-momentum current density, seems capable of *classical Popper-physics* (discrete) 'particle' definition, *regardless of 3-space curvature*. But QFT requires flatness for its 3-space. There is *no* SMU meaning for 'quantum radiation field'.

QFT had become an accurate 'local' approximation for micro to macro spatial scales by the *macro*-scale ages, $\tau/\tau_0$ ~$10^{38}$, when galaxy and star-building commenced--well after the *micro*-scale ages, $\tau/\tau_0$ ~$10^{19}$, when SMU's massive-elementary-particle-building got underway. The Poincaré group and the associated QFT *identical-elementary-particle* micro-macro physics approximation will be addressed by other papers. Appendices here make a start.

QFT's 'parity-reflection' (*not* a Lie-group generator) parallels sign-reversal of SMU's self-adjoint 'chirality'—one of the 3 *central* CL generators. (All 9 generators are 'Noether conserved'). Dirac's writings never mention 'parity' but he proposed a 'doubling' of electron spin through a *velocity direction* (*not* momentum direction) that was either parallel or antiparallel to spin direction.

Dirac's 'relativistic-electron (quantum-physics) wave function' (frustratingly for Dirac, *not* a 'Hilbert vector') satisfied a first-order (Schrödinger) equation of motion via his doubling of 2-valued electron spin. Later this doubling became represented by the notation (0, ½) and (½, 0) for a pair of inequivalent *nonunitary* finite-dimensional SL(2,c) representations. The *three* eigenvalues, 0, ± 1, of SMU's self-adjoint *quc* 'chirality' amount to a 'Dirac tripling' relevant to *all* particles, not only spin-½ fermions.

A pair of main-text 4-vector fiber-bundle *quc* Dirac-*coordinates*, here employed to achieve a retarded Lorentz-tensor-field *classical-cosmological* definition of 'reality', represent the 'Maxwell-Lorentz' group SO(3,1) while *also* providing (½, ½) representation of SL(2,c)$_R$. Classical electromagnetic fields provide (0, 1) and (1, 0) representations of SL(2,c)$_R$, but strikingly-absent from the definition of these fields is any reference to chirality—a term here accorded *cosmological* 'Dirac-tripling' meaning.

Although *no* finite-dimensional representation of *any* 'Lorentz' group is unitary, a *classical* bridge between 'particulate Popper physics' and cosmology is provided by a below-specified Lorentz-tensor 'reality' that does *not* require 3-space to be Euclidean and makes no reference to chirality. *Cosmological absence* of QFT's *quantum-theoretic* (S-matrix) meaning for 'particle' is alleviated by a *classical* cosmological 'Popper' meaning that is based on *charge discreteness* together with Maxwell's equations and an energy-momentum tensor.

Formula (8) here shows how, with GN's *unitary* representation, Dirac *might*, in the spirit of later-appreciated 'supersymmetry', have extended the Hilbert-space-representable U(1) group generated by a self-adjoint operator representing electric charge. The 1-dimensional compact manifold spanned by *chirality*-generated GN Hilbert-vector *argument* displacements covers an interval *twice* that spanned by *charge*-generated displacements of a (complex) Hilbert vector's *phase*.

Formula (8) formalizes the foregoing. *Unitary* SL(2,c) representations remained undiscovered for more than a decade after Dirac's attempt to 'relativize' the electron. Dirac's opinion, either about Milne cosmology or about GN's unitary SL(2,c) representations, is unknown to the author.

The 'exterior' 6-element nonabelian $SL(2,c)_R$ *algebra*, a *subalgebra* of the CL group, [4] defines SMU conserved *momentum* and *angular momentum*—the generators, respectively, of infinitesimal displacements of location and orientation within Milne's negatively-curved metricized 3-space. Milne seems not to have represented either energy or electric charge; he almost certainly did not represent chirality.

The 9-member CL algebra provides a *complete* set of conservation laws to govern SMU evolution of a reality that encompasses Hubble redshift together with *non-particle* extra-galactic dark energy, galactic 'dark matter' *and* QFT's set of macro-scale-observable micro-scale 'identical' *elementary particles within our galaxy*. (Certain *other* galaxies *may* be found to contain an *alternative* set of elementary particles—with QFT's 'left-handed weak-vector bosons' replaced by 'right-handed' counterparts.)

The terms 'particle' and 'identical particles' are FHPP-meaningful at present and recent SMU ages. But Milne-Lorentz *cosmological* symmetry transcends physics by attending to dark matter and dark energy as well as to 'early' SMU history when 3-space curvature was huge on macro scale. [Note: Our adjective 'macro' applies *both* to the (kilometer) 'lab' *spatial* scale at which 'measurements' are performed by 'conscious beings' *and* to the *temporal* scale of those SMU ages when galaxy-clumping began.]

SMU recognizes, while not depending on, *approximate* scale-dependent human-language meaning for 'free-will measurement by an observer'—a macro-scale Galilean notion on which human (Popper) physics is founded. The author believes human language incapable of *'exactly true* sentences'. Any human-language 'truth' we believe a scale-dependent *approximation*, which sometimes may enjoy high accuracy because of ginormous differences between five *different* spatial scales currently recognizable in SMU—GUT scale, micro scale, macro scale, galactic scale and Hubble scale. Human language is *macro*. Far above macro while still far *below* the (present) scale of Hubble is the galactic scale of dark matter. All five scales play roles in the present paper's content, which to the author appears consistent with the belief, by an increasing number of philosophers, that 'free will is a (macro-scale) human illusion'.

Both universe age (approximately measurable by redshift) and the proposed SMU Hamiltonian—Formula (21) below—which 'analytically' generates universe-ray evolution with increasing age, are CL invariant, with the CL algebra (Stone) representable by self-adjoint GN-Hilbert-space (Dirac-momentum) operators. Once again: CL algebra comprises (Noether-conserved) continuous-momentum times age, angular momentum, electric-charge, chirality and energy times age—the latter sextet *all* Dirac-discrete.

**SMU Elemental Constituents**

The unitary CL representation by Formula (8) below has led the author to recognize a set of *SMU elemental constituents*, each here bearing the (pronounceable) name '*quc*' (quantum-universe constituent), which we here suggest compose the *entire* universe in a sense evocative of that accorded by *nuclear* physicists to Gell-Mann's acronym '*quark*'. No *single quc*, individuated by 3 *central integers* specifying its electric charge, its chirality and its energy, is a 'particle'. SMU, nevertheless, we propose to be *completely* '*quc*-composed'.

A subscript $q$, equivalent to a *trio* of integers, $Q_q$, $N_q$ and $M_q$, here distinguishes any *quc* from all other *quc*s. It will below be seen that the *total* number of different *quc*s is 21 $M_{max}$, with $M_{max}$ a ginormous but finite (age-independent) positive integer. Dependence of any *single-quc* wave function on *one* of its 6 'Dirac' *quc* coordinates—that which is 'canonically-conjugate' to $M_q$—'spreads' this *quc* over the *entire* Milne 3-space. Any *individual* particle 'location' similarly is *spread*. *All* 'meaning' is *relative*. Application of *any* element of the CL symmetry group to a (multi-*quc*) SMU ray 'changes *nothing*'.

An SMU ray, at some fixed age, is a *sum* of products with 21 $M_{max}$ factors—each of these a wave function of a *different quc*. An 'elementary particle' is Newton-Maxwell-*theoretically* (Popper) a classical 'clump' of conserved energy, momentum and angular momentum, with an (approximate) 'FHPP mass', a 'spin' and an electric charge equal to some integral multiple of a universal charge unit.

Such 'fixed and settled' Popper reality is later prescribed by (mathematically-defined) *expectations* of certain self-adjoint operators for the SMU ray at the age in question. Any *particle*, whether or not considered 'elementary', associates to *electro-dynamically-correlated* wave functions of *many different* charged *quc*s.

**Two-*Quc* 'Cores' of Massive Elementary Particles**

Within the *individual* terms of an SMU ray's tensor-product summation, a zero-chirality zero-charge 2-*charged-quc* 'core' factor associates to *any* elementary particle *except* photons and neutrinos. A core factor depends on the *quc*-pair's 'relative' coordinate, which later in this paper will be seen essential in a more general context to the SMU Hamiltonian's *potential* energy. A particle's *mass* reflects its core.

Not only does the net charge and chirality of a 'particle-core' *quc*-pair vanish but so does its below-specified baryon number. (A photon or neutrino is a neutral zero-baryon-number but chirality-*bearing quc* pair.) When the elementary particle in question is charged, the charge is provided by a 'valence' *quc*. Also established by valence is particle chirality and baryon number.

What about elementary-particle energy, momentum and angular momentum? A valence *quc* adds its contribution of the foregoing conserved attributes to that of particle core. A Hamiltonian-prescribed *superposition* of products with *variable* values of *total* energy, momentum and angular momentum, as well as different *distributions* thereof between the 3 *quc*s, represents the particle. *Different* energy distributions require *different quc*s to appear within the 2-*quc* or 3-*quc* sets that *quantum*-theoretically represent a *single* elementary particle.

Each *quc* carries a positive *energy* integer with *one* of a ginormous although finite set of possible values, an electric-charge integer with *one* of the 7 values 0, ±1, ±2, ±3 and a 'chirality' integer with *one* of the 3 values, 0, ±1. (*All* particle-building *quc*s have *nonvanishing* electric charge!) The foregoing options became author-appreciated after decades of Occam-guided contemplation that included many discussions with colleagues. Also additively-carried by a particle's constituent-*quc*s are ('familiar') momentum and angular momentum.

**Baryon Number and Chirality**

Temporal microscale *stability* of a particle requires *any particle-composing quc* to have *non-vanishing* electric charge. The electric-charge integer $Q_q$ defines *Quc-q*'s baryon number', $B_q$, which *vanishes* when $Q_q$ takes any one of the three values 0, ±3. If $Q_q$ is either +2 or – 1, $B_q$ is +⅓. If $Q_q$ is either –2 or +1, $B_q$ is –⅓. *Individual quc*s thereby may be categorized as either 'baryonic' or 'nonbaryonic'; total-universe baryon number vanishes. (See Table I in Appendix E.)

Despite absence of 'particle' status for *single quc*s, the 9 *conserved* Dirac-*momentum quc* attributes as well as baryon number are each *additively* manifested by particles 'built' from *quc*s. 'Particles'— (spatial) 'clumps of reality' each with integer net charge and one-third-integer baryon number, odd or even *chirality* and an 'FHPP mass'--are, in a flat 3-space (QFT) approximation, fermions (bosons). *Not* addressed by this paper is the *error* in the S-matrix concept of 'identical' particles.

Particles of *common* FHPP mass, spin, charge, baryon number and momentum but with *differing* energy-*distribution* among constituent *quc*s, are 'physically identical'—FHPP identical--despite *not* being 'cosmologically identical'. (Appendix C.) Pedagogically-useful *quc* segregation, both by angular momentum *and* by *chirality*, into 'fermionic' and 'bosonic' categories is *unaccompanied* by any meaning for '*quc* statistics'.

The physics word 'particle' lacks precise SMU meaning! SMU *classical* cosmology is a 'continuous *onflow* of electro-gravitational reality', with discretely-conserved (because of quantum superselection) electric charge and baryon number and continuously-conserved energy and momentum times age. The author believes the term 'plasma' to be useful, both physically *and* cosmologically, but there is *no* cosmological usefulness for the term 'vacuum'.

[*Energy integer*, whose definition in Reference (4) is repeated below in this paper's main text, was suggested to the author *not* by QFT but by Fermi *momentum* notions useful in *condensed-matter* physics. At present-universe age the difference in energy associating to *successive* SMU energy integers—a difference determined by SMU age--is 'ginormously tiny.']

The SMU Hamiltonian, specified by Formula (21) to be a CL *invariant* (*not* a 4-vector component), comprises a sum of *single-quc* kinetic-energy and *quc-pair* electro-gravitational potential-energy self-adjoint Dirac operators. The *finite* although ginormous set of SMU *quc*s is *age-independent*.

**Cosmological vs. Physical Photons**

Other papers, through 'recent-age' *flat* 3-space (Euclidean) micro-macro *approximations*, will depict as Schrödinger-Dirac *charged-quc composites* not only QFT quarks, charged leptons and W bosons but neutrinos, $Z_0$'s, and Higgs bosons. *Transcending* Euclidean 3-space QFT (whose accuracy derives from *recent* macro-scale hugeness of SMU age) are *net* electrically-neutral but chirality-carrying zero-mass *cosmological* photons ($\gamma_c$)—each a *pair* of nonbaryonic *quc*s *oppositely* charged but of *same nonzero* chirality (total $\gamma_c$ chirality being ±2).

*Equality*, between the value of its chirality and that of twice its *helicity*, accompanies physical-photon *unique* masslessness—*equality* of energy and kinetic energy. The *quc* structure of a propagating $\gamma_c$ might spatio-temporally be described as an 'electro-gravitationally-stabilized *double* helix'—a 'perfect *quc* marriage'—remarkably enjoying the *same* number (6) of 'Dirac degrees of freedom' (*Ddof*) as a *single* chargeless *quc*. (Appendix C) All other *quc* marriages are 'imperfect'.

Two of a $\gamma_c$'s 6 *Ddof* are 'relative' (*internal* and *fluctuating*) coordinates on which a *unique* $\gamma_c$ 'ground-state' internal wave function depends. *Four* of the 6 *Ddof* are *external* 'Dirac momenta' (3-vector momentum plus helicity). Future investigation we expect to expose not only the 'shape' of the $\gamma_c$'s *internal ground state* but *absence* of *other* stable internal states. (The double-helix characterization involves *both* external *and* internal *Ddof* of a *γ$_c$*.)

Pedagogically, the term 'mass' helps to distinguish the 'particle' concept from that of '*quc*'. There is *no* meaning for '*quc* mass'--in contrast to the long-appreciated physics meaning for *zero* photon mass, which Appendix C, via the 2-*quc* *γ$_c$*, accords to FHPP 'identical photons' obeying Bose statistics.

**Miscellany, Perhaps Helpful to Reader Thinking**

Throughout this paper reference to a *single quc* may for convenience of reader thinking be understood as in an 'SMU local frame' whose definition relates to later-defined *quc* 'Dirac coordinates'. FHPP meaning for local frame relates to Milne's celebrated 'cosmological principle'. In local frame, present-universe cosmic background radiation (CMB) is isotropic. (QFT's meaning for 'lab frame' *roughly* matches—with error $\sim 10^{-3}$--that of SMU local frame!) In local frame, time *change* and age *change* are equal—'time interval since big bang' being equal to $\tau-\tau_0$.

'Local frame' associates to the CL-*invariant* meaning of '*quc* energy'. Wherever the latter term here is used it refers to the *quc*'s energy in a space where a (*3-dimensional*) spherically-*symmetric* big bang occurred'—*all* clumping of spatial energy being a consequence of *crd after* universe birth. (Appendix A) Because of clumpings developed *before* ('recent') CMB decoupling from atoms (at $\tau/\tau_0 \sim 10^{57}$), the local frame is only *approximately* establishable by CMB astronomy.

The *quc*-composed Schrödinger-Milne universe may *not* be described as 'composed of elementary particles'. *Additivity* of *quc* 'Dirac momenta' nevertheless allows an 'elementary-particle *set*' of 2 or 3 (*different*) charged *quc*s to represent the CL symmetry group through a Dirac-momentum *unirrep csco*--a complete set of 7 *commuting* self-adjoint operators that adjoins two CL Casimirs to the *direction* (2 operators) of (*exterior*) conserved momentum and the 3 central momenta. SMU's *multi-quc* '*external*-momentum csco' parallels the 'asymptotic Hilbert space' of the particle-physics S matrix. Appendix B addresses the single-*quc* 'Dirac-momentum' csco.

The SMU Hamiltonian--to which this paper's main text leads--is expressed through Dirac *quc* coordinates *and* Dirac *quc* momenta, but *not* through particles. The author nevertheless expects SMU's Hamiltonian (eventually) to *explain* (approximately) the observed values of elementary-particle masses and other arbitrary QFT parameters. (Appendix E)

Each of 9 *conserved* elementary-particle attributes is the *sum* of (corresponding) constituent-*quc* attributes. QFT elementary-particle 'asymptotic Hilbert space' enjoys useful approximate (flat 3-space) S-matrix meaning because the macro scale of a human laboratory, although huge on micro scale, is tiny on Hubble scale. Sums of *quc* (Dirac) momenta approximate particle (S-matrix) momenta. No such feature attaches to (non-conserved) *quc* (Dirac) coordinates.

Even with quantum fluctuation of the energy integer, $M_q$ (essential to single-*quc* identification), it is useful to think of any QFT elementary particle as a *macro*-stable ('Popper-measurement-accessible') potential-energy-induced marriage (*after* universe birth and of GUT or *micro* internal spatial scale) between 2 or 3 electrically-charged chirality-carrying *quc*s whose wave functions at SMU birth were *uncorrelated*. (Appendix A)

An 'early' such marriage--at GUT or micro-scale ages--might *dynamically* be described as Hamiltonian-generated 'collapse' of a product of *uncorrelated* charged single-*quc* wave functions into a *macro*-temporally-stable *micro* or GUT-sized 'object' wave function. (The photon is the *only* 'elementary' particle where, because of *GUT-scale* internal extension, *gravity* significantly contributes to stability.) In any present-day S-matrix 'connected part', it is electromagnetic potential energy that causes charged married *quc*s to 'change partners'.

Dark matter comprises *gravitationally*-sustained *galactic-scale* 'colonies' of *individually-chargeless* bachelor *quc*s. (A dark-matter bachelor-*quc* wave function spatially *spans* its galaxy.) *Inter*-galactic universe-spanning bachelor *quc*s, both charged and uncharged but without previous 'marriage history' because of energy too high for electromagnetic macro-clumping, constitute 'dark energy'.

Particle *mass*, a physics word without (onflow) cosmological meaning, reflects Hilbert-space collaboration between the *discrete ca* trio (electric charge, chirality, energy) and *continuous* zitterbewegung (*zbw*). The latter (Schrödinger-coined) term refers to *fluctuation* of lightlike velocity *direction*--a *non*-conserved Dirac *coordinate*--at *fixed* conserved Dirac *momenta*. As above noted, the CL *unirrep csco* provides a Hilbert-space FHPP elementary-particle-momentum basis that, although without particle mass among its labels, resembles 'in'→'out' S-matrix 'asymptotic Hilbert space'.

*Physics-foundational*, however, are conserved (and with *commuting* components) *energy-momentum 4-vector* Poincaré-group generators. Essential SMU roles are played in this paper's main text by two non-conserved positive-4-vector *quc* Dirac *coordinates*, specifying the *quc*'s *spatio-temporal location* and its *lightlike-velocity direction*, but there are *no* SMU *quc-energy-momentum* 4-vectors. SMU momentum-basis portrayal of a photon's external *Ddof* extends to massive elementary particles--via CL-invariant particle *energy* and a *smaller* invariant, although not conserved, momentum *magnitude* ('relativistic kinetic energy'). But *neither* portrayal is 4-vector!

For pedagogical reasons the CL *unirrep* (Dirac momentum basis) is *not* addressed by this paper's main text. Appendix B exposes the *quc* Dirac-*momentum* csco that complements the main-text-explored *quc* Dirac-*coordinate* csco.

QFT is a spatially-restricted scale-dependent *approximation* that accords meaning to 'vacuum', ignores (photon-redshifting) universe expansion and relies on an S matrix with *a priori* elementary-particle masses and electric-charge screening. Electron mass is treated by QFT as *non*-fluctuating and age independent.

*SMU* (approximate) *age-independence* of electron mass we associate to *ongoing* interchange between electron-attached *quc*s and charged intergalactic bachelor (un-clumped 'dark-energy') *quc*s of, on average, slightly *higher* $M_q$ than those of the electron *quc*s they replace. The author counts on eventual verification of such interchange by Hamiltonian-based computation. The 'vacuum' concept is *absent* from SMU, where the set of *quc*s is *fixed*. *Incompleteness* of electric-charge screening is of *major* SMU importance.

*Any quc* subset represents CL. The *total* set of universe-comprising *quc*s, although ginormous, is finite and *constant* (*τ*-independent). *Quc*s are *never* created or annihilated. We repeat: Despite each of 8 CL generators--all *except* chirality--representing a conserved *quc* attribute to which a physics-familiar name may be attached, while all 9 *ca* are particle-carried, *no quc* may ever *individually* be called 'particle'.

Six conserved 'exterior' attributes of a *quc*—its momentum and its angular momentum—may change from *crd* interaction with other *quc*s as universe-age increases. Three integer-specified conserved 'central' attributes--electric charge, chirality and energy times age—remain unchanged. *Continuous* momentum-basis variability of a *quc*'s *momentum-magnitude* times age, while its energy integer remains fixed, renders impossible any definition of 'single-*quc* mass'.

O*bjective* reality—*spatially-localized* and *temporally-stable* current density--involves at the very least *two quc*s. A single *quc* cannot represent an 'object'—the definition of which requires a self-sustaining relationship between *different quc*s. Dark matter, despite its gravitationally-sustained localization being at galactic scale, *is* 'objectively real'. A galaxy, although not built entirely from 'particles', *may* astronomically be described (approximately) as a 'spatially-localized object'.

Any *future* SMU (of age greater than present age) is determined by the *quantum* state of SMU *now* (as the reader is seeing this sentence). SMU's *quc* foundation transcends measurement-based physics. A self-adjoint Hamiltonian operator is a Schrödinger-Dirac-imitating sum of single-*quc* kinetic-energy and *quc*-pair potential-energy operators.

Each *quc* carries a discrete energy that *permanently* remains positive despite perpetual *decrease* by its inverse proportionality to *positive-increasing* universe age. Accompanying perpetual photon redshift is ongoing (*never complete*) flattening of hyperbolic 3-space. [4] These SMU features (in profound contrast to general relativity) involve *no* reference to 'local' energy *density*!

Current densities of (classical) energy-momentum and electric charge—specified by *expectations* of E-G field-operators—define at every age an SMU 'reality'. Only a *portion* of this reality, nevertheless, is 'objective'—expressible through the 'stable object' notion (which includes a 'galaxy of stars with attached dark matter'). *Most* of the *present* universe's energy remains non-objective. Bohm hidden-reality comprises both dark energy and, *wrt* 'Copenhagen (S-matrix) quantum physics', dark matter. Disregard of SMU's hidden reality has required probabilistic interpretation for 'Copenhagen' quantum theory.

At any universe age, $\tau \geq \tau_0$, an SMU Hilbert-space ray, 'regularly' representing CL, is a complex normed multiply-differentiable function of the 6 *Dirac coordinates* of *each* member of a ginormous but finite age-independent set of *quc*s. We shall see how the six coordinates of any *quc*, equivalent to 3 complex coordinates, specify via a complex unimodular $2 \times 2$ *coordinate* matrix the *quc*'s location within a 6-dimensional manifold. A *unit* such matrix locates the *quc* at an SMU 'oriented center'. *Any* 'exterior *quc* location'—later shown to include a 5-dimensional fiber bundle--is rendered 'central' by that right-SL(2,c) transformation (an exterior element of the CL group) which transforms this *quc*'s Dirac coordinate to a unit matrix.

Fixed Popper reality--current densities of energy, momentum, angular momentum and electric charge (*not* chirality)--is prescribed by the Dalembertian of a 13-component ($2^2+3^2$) retarded *classical* electro-gravitational (E-G) Lorentz-tensor field. The present paper's main text specifies this field by SMU-ray *expectations* of *self-adjoint* Lorentz-tensor field (*not* radiation-field) operators.

At SMU birth *all* reality was hidden and *full* reality *fails* (at *any* age) to specify the SMU *ray* that is Schrödinger-determined by the 'immediately-preceding' ray. Present reality *fails* to determine future reality! Hilbert space with Schrödinger Hamiltonian dynamics is SMU essential.

**Unification**

This paper 'unifies' gravity and electromagnetism by 'bundling' *classical* Newton-Maxwell (*G-c*), *quantum* Planck-Schrödinger-Dirac ($\hbar$) and *classical* Hubble-Milne ($\tau$). The foregoing brackets associate to natural philosophers symbols for 4 positive dimension-ful real parameters--3 constant (*G*, *c*, $\hbar$) and one perpetually increasing ($\tau$)--that underpin the present paper.

*Any* objective reality, such as a photon, a proton, a molecule, a planet, a star or a galaxy, associates to *exceptional* temporally-stable spatially-localized *multi-quc* wave functions where the 'expansion' tendency of *positive quc* kinetic energy--to *increase* spatial separation between *different qucs*—is opposable by *negative* gravitational and (or) electromagnetic *potential* energy that tends to *decrease* separation.

'Strong interactions'('nuclear forces') arise from the Formula (21) Hamiltonian's kinetic plus *electromagnetic* potential energy, applied to systems with baryon-number-carrying valence *quc*s. *Gravitational* potential energy, *together with* electromagnetic, we believe essential at GUT scale to the *photon double helix* as well as, at ginormously larger scales, to planets, stars and galaxies.

**'Quality' 3-Space; Natural Units**

SMU displays a 3-dimensional 'quality space'. The *central* SMU 'momentum dimensionalities'—those of energy, electric charge and chirality--span *all* universe dimensionality. SMU chirality *shares* the dimensionality of angular momentum. SMU associates $G$ (Newton) to energy, $c$ (Maxwell) to electric charge and $\hbar$ (Planck) to chirality.

*Dimensionless* non-conserved *quc* attributes prominently include (as will later be seen) 2-dimensional positive-lightlike *velocity direction*--an SMU 'Dirac *quc* coordinate'. Mathematics distinction between number-theoretic 'topology' and analysis-theoretic 'geometry' provides some analog to the distinction between 'dimensionless' and 'dimension-ful' *quc* attributes, but neither mathematics nor theoretical physics has so far dignified through a symbol the 3-dimensional 'quality space' displayed by the universe. Such a symbol would 'legitimatize' SMU's Lie symmetry-group *center*--currently puzzling to mathematicians as well as physicists.

Quality-space 3-dimensionality dovetails with the *trio* of *independent* universal dimension-ful constants with *independent* dimensionalities. To these constants the symbols $G$, $c$ and $\hbar$ have become attached. Occam-inclined natural philosophers expect *no* universal *dimension-ful* constant (*udc*) *beyond* this trio.

Notational economy has long been appreciated for units that assign the value 1 to each member of the *udc* trio. This paper will henceforth employ such units, thereby attaching unambiguous numerical value to the dimension-ful SMU age symbol $\tau$ (present-universe age in 'natural' units being $\sim 10^{60}$), to the 'big-bang' birth age $\tau_0$ (1 or ten to power zero), and to dimension-ful particle-physics parameters such as Higgs mass ($\sim 10^{-17}$) and electron mass ($\sim 10^{-22}$).

Values for *all* the many arbitrary QFT parameters we expect to be shown SMU-Hamiltonian-determined by the (macro) temporal-stability spatial-localizability requirement implicit in objectivity. [Our guess for 'diameter of photon double helix'—a notion unrecognized by QFT--locates in the neighborhood of QFT's 'GUT' spacetime scale-- $\sim 10^4$ in natural units (Appendix C).]

**Mysterious Physics-Enabling Macro-Scale; Avogadro Number; Feynman's Perturbative S Matrix**

This paper assigns 'foundational' status neither to the micro spatial scale, $\sim 10^{19}$, set (inversely) by elementary-particle masses nor to the much larger ($\sim 10^{38}$) 'macro' spatial scale that characterizes at once the (kilometer) scale of human 'laboratories' and the Schwartzschild radius of stellar *mass*. These scales we portray here as *crd outcomes*. Far below micro spatial scale is SMU-foundational GUT scale—that we associate to the electric-charge unit $g$. Far above macro are galactic and (present) Hubble scales. 'Macro' locates in the 'logarithmic middle'. Two sections below we mention a huge Mersenne prime that, via $M_{max}$, might relate to 'macro'.

QFT's spatial parameters are 'micro'--in a neighborhood *below* macro by a factor *larger* than the cube root of Avogadro's mysteriously-huge number. Long appreciated is dependence of physical chemistry, statistical mechanics and condensed-matter physics on the latter's largeness.

Macro scale understood as 'lab scale'--the 'scale of measurement'--is recognized by the QFT S-matrix as well as by Avogadro-number-based atomic and molecular physics, statistical mechanics and condensed-matter physics. How does QFT recognize macro scale? Within each denominator of Feynman's perturbative series for an S-matrix element there appears the symbol ε—representing an energy which, although 'vanishingly-small' for S-matrix purposes, cannot be ignored. S-matrix *definition* regards 'macro time' as 'ginormously larger' than the times associating to elementary-particle masses.

The *inverse* of Feynman's ε may be regarded QFT's *definition* of 'macro' time scale--huge for particle-physics purposes, as well as for those of any science relying on Avogadro-number hugeness, while 'tiny' compared to galactic scale and yet tinier when compared to that of Hubble. 'Macro scale' includes that extremely narrow scale range where 'free-will measurement by conscious life' enjoys meaning. Alfred North Whitehead acknowledged mysterious macro scale not by a number but through use of the term, 'God'. Schwartzschild was thinking 'stellar black-hole radius' when recognizing macro scale.

**Age-Independent Huge Finite Set of *Different Quc*s**

SMU is populated by a finite set of *different quc*s—distinguished by an electric-charge integer, $Q_q$, a chirality integer, $N_q$, *and* a positive energy integer, $M_q$. Each of the corresponding attributes enjoys a separate dimensionful unit. QMU has exactly *one quc* for any $Q_q$, $N_q$, $M_q$ *integer-trio*, with $Q_q$ allowed the 7 values 0, ±1, ±2, ±3 and $N_q$ allowed the 3 values 0, ±1 while $M_q$ is allowed $M_{max}$ possible values, 1, 2, ... $M_{max}$. The total number of SMU *quc*s is thus 21 $M_{max}$. Our choice of allowed values for $Q_q$ and $N_q$ has been influenced by Occam, by Mersenne and by QFT's set of elementary particles. (Appendix E.) For present-paper purposes we elect to leave still unspecified the ginormous value of $M_{max}$.

Any DMU ray is a sum of products of 21 $M_{max}$ single-*quc* normed functions, any *quc* appearing exactly *once* in each such 'tensor' product. It will below be seen that GN unitary Hilbert-space representation of SL(2,c) requires each single-*quc* Dirac-coordinate-basis wave function to depend on that *quc*'s location within a noncompact *4-dimensional* manifold--*not* a spacetime manifold but a product of two complex-variable-coordinated manifolds.

An approximately temporally-stable multi-*quc* although single-*particle* wave function, such as that of an electron, *correlates* constituent-*quc* energy-integers with constituent-*quc* chiralities and charges. In 'reactions' that annihilate this particle while creating other particles, *quc* 'creation' or 'annihilation' never occurs. Instead there is *quc* 'reallocation'.

Sub-product clusters of 2 or 3 *quc*s represent elementary particles of *sharply* (integer)-specified electric charge and baryon number. *Chirality evenness or oddness* of any particle is unambiguous. *Always*, within any elementary-particle wave function there is superposition of different energies whose spacing, $(2\tau)^{-1}$, is presently smaller than any 'soft-photon' energy by a factor of order $10^{22}$.

The author anticipates eventual number-theoretic specification of $M_{max}$, the maximum *quc* local-frame energy in units of $1/2\tau$. Perhaps $M_{max}$ is a *ginormous* Mersenne prime. $M_{max}$ is one of several SMU 'foundational numbers'. Another is the dimensionless coefficient *g*, related to the 'fine-structure constant' of particle physics, that determines GUT scale and appears below in our formulas for electric-current density and Hamiltonian electromagnetic potential energy. The 'large' Mersenne prime $2^7-1=127$ has often been conjectured to set the value of $g^{-2}$. Our number-theoretic Occam-choice for a third foundational parameter--the SMU birth age $\tau_0$ in 'natural' units--is 1. The huge Mersenne prime, $2^{127}-1$, may relate to $M_{max}$ and to mysterious macro scale.

As age $\tau$ ($> \tau_0$) continuously increases, the value of every *quc*'s energy--some fixed-integer multiple of $1/2\tau$--decreases in inverse proportion to age: *Milne redshift*. The value of (continuous, positive, dimensionful and global) *age* establishes an SMU scale (presently 'Hubble') not only for *quc* total and kinetic energies [see Formula (21)], but for *quc* momentum (main-text section on EL Casimirs) *and* for distance (at same age) between different *quc* locations in the hyperbolic Milne 3-space [3] whose Riemann (negative) curvature is $\tau$-determined. [4]

Classical (although *quc*-sourced) retarded E-G tensor fields whose Dalembertians prescribe SMU reality, will be seen below to have 13 components. The electromagnetic and gravitational fields sourced by *Quc q* are proportional, respectively, to $gQ_q$ and to $M_q/2\tau$. The former is the *quc*'s electric charge while the latter is its energy. *Total* universe *energy* at age $\tau$ is, by a simple computation, $21(M_{max}/2)^2 \tau^{-1}$, while total charge and total chirality vanish, together with total baryon number.

**Unitary Hilbert-Space Dirac-Coordinate Representation of 9-Parameter CL**

We now reproduce, with minor notation adjustment, certain formalism from Reference (4). An SMU ray at Age $\tau$ is a sum of ('tensor') products, each with $21M_{max}$ factors, of single-*quc* Hilbert vectors. In the 'Dirac-coordinate' Hilbert-space basis, each of the latter is a normed complex differentiable function, $\psi_q^\tau(\boldsymbol{a}_q)$, of *Quc q*'s location $\boldsymbol{a}_q$ in a 6-dimensional manifold.

The symbol $\boldsymbol{a}_q$ denotes the spectra of a *complete set of 6 commuting self-adjoint operators*—a 'Dirac-*coordinate*' *csco* that complements a 6-element, *Casimir*-based, 'Dirac-*momentum*' *csco* which our main text ignores, apart from references to CL Casimirs. Appendix B attends to the momentum basis.

Reference (4) has shown how $\boldsymbol{a}_q$ comprises a 3-dimensional *metricized base-space* location *and* a (non-metricized) *fiber-space* location whose dimensionality is 2 + 1 = 3. The coordinate of the 1-dimensional fiber subpace will be seen Dirac-conjugate to chirality. The *complete* 6-dimensional '*quc*-locating' coordinate amounts to a *2×2 complex unimodular matrix*. Henceforth in this paper's main text *any* boldface symbol is to be understood as denoting a 2×2 matrix.

The single-*quc* Dirac-coordinate-basis wave function, $\psi_q^\tau(\boldsymbol{a}_q)$, unitarily representing CL at each age $\tau \geq \tau_0$ [see Formula (8)], is a Hilbert vector with the invariant (finite) norm,

$$\int d\boldsymbol{a}_q \left| \psi_q^\tau(\boldsymbol{a}_q) \right|^2 . \qquad (1)$$

The CL-invariant 6-dimensional volume element (Haar measure) $d\boldsymbol{a}_q$ we below express through a *trio* of *complex* Dirac coordinates equivalent to the matrix $\boldsymbol{a}_q$.

Because the present section and that following refer to a *single quc* and a *single* age, we shall in these sections omit both the superscript $\tau$ *and* the subscript $q$. Also ignored, *except* in *Eq*. (8), is the charge integer $Q_q$; U(1) transformation does *not* affect the wave function's (Dirac-coordinate) *argument*-- merely shifting complex-wave-function *phase* (in *any* basis) by an increment proportional to $Q_q$.

The unimodular 2×2 complex (Dirac) *quc*-coordinate matrix $\boldsymbol{a}$ is equivalent to *three* complex variables: *s*, *y*, *z* (six real variables), according to the following product of three unimodular 2×2 matrices, each of which coordinates the manifold of a 2-parameter abelian CL subgroup :

$$\boldsymbol{a}(s, y, z) = exp(-\boldsymbol{\sigma}_3 s) \times exp(\boldsymbol{\sigma}_+ y) \times exp(\boldsymbol{\sigma}_- z). \qquad (2)$$

The *s* subgroup lies within the CL (diagonal-matrix) center. The complex variables *s* and *y together* coordinate the manifold of a 4-parameter nonabelian CL subgroup. The latter feature is essential both to GN's unitary transformation between Dirac-coordinate and Dirac-momentum bases [2] and to coordination of base and fiber spaces.

The 2×2 real-matrix pair $\boldsymbol{\sigma}_\pm$ is defined as $\frac{1}{2}(\boldsymbol{\sigma}_1 \pm i\boldsymbol{\sigma}_2)$. The (familiar to physicists) Pauli-matrix symbols $\boldsymbol{\sigma}_3$ and $\boldsymbol{\sigma}_1$ represent hermitian *real* self-inverse traceless 2×2 matrices with determinant –1, $\boldsymbol{\sigma}_3$ being diagonal and $\boldsymbol{\sigma}_1$ off-diagonal, while the symbol $\boldsymbol{\sigma}_2$ represents an *imaginary* such matrix equal to $-i\boldsymbol{\sigma}_3 \boldsymbol{\sigma}_1$. The 6-dimensional Haar measure,

$$d\mathbf{a} = ds\, dy\, dz, \qquad (3)$$

is invariant under $\mathbf{a} \to \mathbf{a}^\Gamma \equiv \mathbf{a}\Gamma^{-1}$, with $\Gamma$ a 2×2 unimodular complex matrix representing a *right* SL(2,c) transformation of the coordinate $\mathbf{a}$. The measure (3) is *also* invariant under analogous *left* transformation. Any 'volume-element' symbol $d\xi$ in (3), with $\xi$ complex, means $d\,Re\,\xi \times d\,Im\,\xi$.

The Hilbert-vector norm (and inner-product)-defining integration (1) is, *wrt Im s*, over *any* continuous $2\pi$ interval of *Im s*. Interpreting $2\tau Re\, s$ as *periodic* '*quc* time' we shall below shrink the Hilbert space so that *Re s* and *Im s* enjoy similar status in vector-norm (and inner-product) regular-basis definition. *Full* real lines for *Re y*, *Im y*, *Re z* and *Im z* remain spanned by the shrunken space's vector-norm definition.

A transformation specified by the 2×2 complex unimodular *right*-acting matrix $\Gamma$ is *unitarily* Hilbert-space represented by

$$\Psi(\mathbf{a}) \to \Psi(\mathbf{a}\Gamma^{-1}). \qquad (4)$$

Straightforward calculation shows $\mathbf{a}\Gamma^{-1}$ to be equivalent to

$$z^\Gamma = (\Gamma_{22}z - \Gamma_{21})/(\Gamma_{11} - \Gamma_{12}z), \qquad (5)$$

$$y^\Gamma = (\Gamma_{11} - \Gamma_{12}z)[(\Gamma_{11} - \Gamma_{12}z)y - \Gamma_{12}], \qquad (6)$$

$$s^\Gamma = s + \ln(\Gamma_{11} - \Gamma_{12}z). \qquad (7)$$

Under the 9-parameter CL symmetry group the 2-dimensional volume element *ds* within the Haar measure (3) is seen from (7) to be invariant. *Also* invariant is the 4-dimensional volume element *dy dz*.

We now explicitly display CL representation by the single-*quc* Dirac-coordinate Hilbert-space basis. A CL element (a location within the 9-dimensional CL manifold) is specified by a U(1)-representing angle, $\omega$, with $0 \leq \omega < 2\pi$ (1 parameter), by a *left-acting* SL(2,c,D) complex *argument* displacement $s \to s + \Delta$ (2-parameter) and, finally, by the Formula (4) SL(2.c)$_R$-representing argument displacement (6-parameter). Under a so-specified (9-parameter) CL element,

$$\Psi_Q(s, y, z)_{(\omega, \Delta, \Gamma)} \to e^{iQ\omega}\, \Psi(s^\Gamma + \Delta, y^\Gamma, z^\Gamma). \qquad (8)$$

*Essential* is commutativity of the *two* s-displacements. The 1-dimensional Haar-measure volume elements $d(Re\, s)$ and $d(Im\, s)$ are *separately* CL invariant, together with the (4-dimensional) *dy dz* volume element.

**Periodicity in '*Quc*-Time'--a Hilbert-Space Reduction**

Displacement in the coordinate *Re s*, at fixed *Im s*, *y*, *z* and $\tau$, displaces what we choose to call '*quc* local time' at fixed values of global age and the *quc*'s 5 other coordinates. *Quc* energy--a self-adjoint Hilbert-space operator representing a member of the CL *center* subalgebra and the source of SMU gravity--is *canonically-conjugate* in Dirac sense to $2\tau Re\, s$.

Although positive lightlikeness of a *quc* velocity 4-vector (defined in the following section) invites confusion between 'temporal' and 'spatial' *quc* displacement, the group algebra unambiguously distinguishes CL-invariant *quc energy* from any non-invariant *quc-momentum* 3-vector component of a right 6-vector—an algebra member that generates infinitesimal *quc* spatial displacement (at fixed age) in some (arbitrarily-specifiable) direction through curved metricized 3-dimensional base-space. [4]

[Because the infinitesimal-displacement direction is specified in some *fixed* right-Lorentz frame, whereas a geodesic follows a curved path requiring *parallel transport* of direction-defining axes, the later-defined invariant self-adjoint *quc* kinetic energy—a function of *Casimir* geodesic-associated *second* derivatives—is *not* proportional to the 3-vector inner product with itself of *quc*-momentum.]

Already noted has been the explicit indication by Formula (7) that (fixed-*τ*,*y*,*z*) displacements in *s* are right-Lorentz invariant (both real and imaginary parts). They further are invariant under the 3-parameter symmetry (central) subgroup (with energy, chirality and electric charge as generators) that defines *quc* type, despite failure to be invariant under the full (6-parameter) *left*-Lorentz group. A *quc* Hilbert-space *shrinkage* that requires ray *periodicity* in regular-basis dependence on *Re s* (location of a '*quc*-timepiece hand') maintains *quc* capacity to represent 9-parameter CL.

We therefore diminish *quc* Hilbert space by the Dirac-coordinate-basis *periodicity* constraint,

$$\Psi(s, y, z) = \Psi(s + 2\pi, y, z), \tag{9}$$

modifying the Hilbert-vector norm to integration in (1) over any (single, continuous) $2\pi$ interval of *Re s*, as well as of *Im s*. The constraint (9) specifies *integer* eigenvalues for the self-adjoint operator that is canonically-conjugate to *Re s*. (The *quc*-energy operator has eigenvalues $M/2\tau$.) We further reduce the Hilbert space by requiring energy to be *positive* and upper bounded (*M* a finite *positive* integer).

For each *quc* there are 9 'conserved-momentum' self-adjoint operators, although they do not all commute with each other. A 6-element Dirac-*momentum* csco that includes the operators *M* and *N* is identified in Appendix B with an 'irreducible' CL representation. The operators *M* and *N* are, respectively, canonically-conjugate to the operators *Re s* and *Im s*.

**'Dirac-Coordinate' 4-Vector Operators that Locate a *Quc* in a 5-Dimensional Fiber Bundle**

Reference (4) defines a (classical, 'exterior') positive 4-vector as a (4-parameter) *positive-hermitian* 2×2 matrix that transforms under an exterior Lorentz transformation $\Gamma$ through right multiplication by $\Gamma^{-1}$ *and* left multiplication by the hermitian conjugate of $\Gamma^{-1}$. A 4-vector's invariant 'squared magnitude' is the hermitian-matrix's determinant. (The timelike component is half the hermitian-matrix trace.) Timelike, lightlike and spacelike 4-vector matrices have, respectively, positive, zero and negative determinants.

A *pair* of commuting (exterior) 4-vector Dirac-*coordinate* self-adjoint *operators*, one positive-timelike and one positive lightlike, are equivalent to a quintet of real *quc* coordinates that specifies the following *Quc-q* 2×2 unimodular matrix, [4]

$$\boldsymbol{b}_q \equiv exp\,(\boldsymbol{\sigma}_3 s_q) \times \boldsymbol{a}_q \tag{10}$$

$$= exp(\boldsymbol{\sigma}_+ y_q) \times exp(\boldsymbol{\sigma}_- z_q). \tag{11}$$

Henceforth *any* symbol with a *q* subscript, whether attached to a unit 3-vector (see below), a 2×2 matrix (boldface indicated, as in the foregoing formulas), to a Lorentz tensor with 4-valued indices *or* to an EL Casimir, is to be understood as a self-adjoint (Dirac) *Quc-q operator* on the SMU Hilbert space.

The *left* multiplication (10) of the *quc*-coordinate matrix $\boldsymbol{a}_q$ by the *diagonal unitary* unimodular matrix, $exp\,(\boldsymbol{\sigma}_3 s_q)$, has *deleted* the coordinates $s_q$ from the coordinate-sextet $s_q$, $y_q$, $z_q$ (a csco). The eliminated coordinate pair (understood as self-adjoint operators), are canonically-conjugate in Dirac sense to those conserved integer-eigenvalued *Quc-q momenta* (elements of the CL algebra) which we have called 'energy' and 'chirality'.

*Quc-q*'s 'fiber-bundle location', either in 3-dimensional metricized *base space* or in a 2-dimensional (unmetricized) 'velocity-direction' *fiber space*, *fails* to depend on $s_q$. We now show that base 3-space location and velocity-direction fiber 2-space location are equivalent to the 4 coordinates, $y_q$, $z_q$— the latter symbols representing a set of 4 commuting self-adjoint Dirac-coordinate operators collectively representable by the (single) 2×2 unimodular-matrix symbol $\boldsymbol{b}_q$.

*Either* coordination associates in Dirac sense to a *quartet* of commuting self-adjoint *coordinate* operators on a single-*quc* Hilbert space. Throughout the remainder of this paper's main text, *all quc*- *coordinate* symbols are to be understood as referring to self-adjoint Dirac operators, either on single-*quc* or multi-*quc* Hilbert spaces. Such operators commute *neither* with CL's *momentum* algebra *nor* with Casimir quadratic functions of that algebra.

Formulas (2) and (9) together expose as *positive-hermitian* the unimodular (*single-quc*) 2×2 *coordinate* matrix,

$$\boldsymbol{B}_q \equiv \boldsymbol{b}_q^\dagger \boldsymbol{b}_q \tag{12}$$

$$= exp(-\beta_q \boldsymbol{\sigma} \cdot n_q), \tag{13}$$

the (non-matrix) symbol $\beta_q$ in (13) denoting a rotationally-invariant non-negative continuous-spectrum self-adjoint operator while the (non-matrix) symbol $n_q$ denotes a *unit 3-vector* self-adjoint operator that commutes with $\beta_q$. The symbol $\boldsymbol{\sigma} \cdot n_q$ denotes the inner product of *two* 3-vectors—*one* a Hilbert-space operator and the *other* a 2×2 hermitian matrix. [4] The hermitian unimodular 2×2 matrix (*and* self-adjoint Dirac-operator) symbol $\boldsymbol{B}_q$ denotes a positive-timelike dimensionless (exterior) 4-vector of unit 'Minkowski magnitude'.

Through the dimension-ful positive-real factor $\tau$ (age) the positive-timelike 4-vector (operator) symbol $\boldsymbol{x}_q \equiv \tau \boldsymbol{B}_q$ locates the *quc* within (non-Riemannian) Milne (Minkowski) spacetime—prescribing (in Dirac-operator sense) its displacement from the vertex of the universal forward lightcone. In a (to physicists) more familiar notation, the 4 (operator) components of $\boldsymbol{x}_q$ are $\tau \cosh \beta_q$, $\tau n_q \sinh \beta_q$.

[Warning to physicists familiar with the Copenhagen statistical interpretation of Dirac quantum theory: SMU *reality* does *not* associate to the expectation of *single-quc* self-adjoint operators, such as $\boldsymbol{x}_q$--attaching *rather* to the *expectations* of self-adjoint E-G *potential*-field operators on the multi-*quc* Hilbert space—operators defined in the section which follows. The foregoing operators do *not* include zero-Dalembertian 'radiation' fields of the kind that QFT associates to its elementary particles.]

Complementing dimensionless $\boldsymbol{B}_q$, which coordinates a *quc*-fiber-bundle's metricized base space, is a second dimensionless positive 4-vector—this one lightlike--to be denoted by the symbol $\boldsymbol{v}_q$ and coordinating a 2-dimensional unmetricized fiber space. The *pair* of 4-vectors, $\boldsymbol{B}_q$, $\boldsymbol{v}_q$, is equivalent to $\boldsymbol{b}_q$ and thereby to $y_q$, $z_q$.

The latter equivalence is below exhibited via invariant 4-vector inner products. The inner product of two 4-vectors will be denoted by the symbol •. The inner product of two *positive* 4-vectors is non-negative. [Because the inner product of *any* two *right* (exterior) 4-vector operators may be shown equal to the inner product of a unitarily-equivalent *left* 4-vector pair, [4] *either* product is invariant under the 12-parameter group $SL(2c)_L \times SL(2c)_R$ and thereby is CL invariant.]

The *quc*-velocity-direction positive-lightlike 4-vector self-adjoint Dirac-coordinate operator, $\boldsymbol{v}_q$, is defined to be the dimensionless *zero-determinant* positive-hermitian matrix

$$\boldsymbol{v}_q \equiv \boldsymbol{b}_q^\dagger (\boldsymbol{\sigma_0} - \boldsymbol{\sigma_3}) \boldsymbol{b}_q, \tag{14}$$

the symbol $\boldsymbol{\sigma_0}$ here denoting the *unit* 2×2 matrix. Equivalence of the coordinate matrix $\boldsymbol{b}_q$ to the dimensionless positive 4-vector *pair* $\boldsymbol{B}_q$, $\boldsymbol{v}_q$ follows from the inner-product trio, $\boldsymbol{B}_q \cdot \boldsymbol{v}_q = 1$, $\boldsymbol{B}_q \cdot \boldsymbol{B}_q = 1$ and $\boldsymbol{v}_q \cdot \boldsymbol{v}_q = 0$, deducible by going to the special ('local') frame where $\boldsymbol{B}_q = \boldsymbol{\sigma_0}$ (i.e., where $\beta_q = 0$).

In the physicist-familiar 4-component notation, the 4-vector self-adjoint operator $\boldsymbol{v}_q$ is equal to

$$(1, u_q) / (\cosh \beta_q - u_q \cdot n_q \sinh \beta_q), \tag{15}$$

$u_q$ a $z_q$-equivalent (2-dimensional, see Appendix B) unit 3-vector (a self-adjoint Dirac-coordinate operator accompanying the unit 3-vector $n_q$ but independent thereof) that admits the name, 'direction of *quc* lightlike-velocity'. A *quc* fiber bundle is thus coordinated by the (3-dimensional) base-space coordinates, $\beta_q$, $n_q$, together with the (2-dimensional) velocity-direction fiber coordinate $u_q$. An equivalent Dirac-coordinate set is $y_q$, $z_q$.

The positive-lightlike *quc* 4-velocity $v_q$ will be seen in the following section, together with $Q_q$ and $M_q$, 'almost completely' to specify 'electromagnetism-gravity from *Quc-q source*'.

## *Classical* Retarded E-G (*Non*-Radiation) Fields

As emphasized in the preceding section, throughout this paper's remaining main text *any* symbol with a *quc*-designating subscript is to be understood as representing a self-adjoint *age-independent* Dirac operator. The *only* age-*dependent* self-adjoint SMU operator is the Hamiltonian. (SMU does not admit the 'Heisenberg picture'. *No* operator represents '*quc* acceleration'.)

The present section deals with *classical retarded* electromagnetic and gravitational Lorentz-tensor fields (*not* zero-Dalembertian 'radiation fields') whose Dalembertian prescribes SMU's 'reality'. At any age $\tau \geq \tau_0$, such fields are prescribed by *expectations*, with respect to the SMU ray at that age, of certain self-adjoint *retarded* Hilbert-space *operators* which sum over all *quc sources* of the field in question. These retarded-field tensor operators we now define through the previous section's pair of 4-vector single-*quc* Dirac-*coordinate* operators.

*Four* electromagnetic SMU field components are complemented by *nine* gravitational-field components. The preceding section's *quc-fiber-bundle* coordinating 2-parameter positive lightlike *quc*-velocity 4-vector and 3-parameter *quc* spacetime-location 4-vector, together with *quc* electric charge and energy, 'co-variantly' prescribe a '*quc*-source' for 13 E-G retarded-field operators. *Positivity* of *quc*-velocity 4-vector associates to the source's *retarded* nature; this 4-vector also prescribes the 'direction' of the generated tensor field. The source-*location* 4-vector, through the denominator of the Lienard-Wiechert (*LW*) formula, joins *quc* charge and energy to determine *magnitude* of the *quc*-generated field.

We define retarded electro-gravitational field operators--*not* 'radiation' quantum fields--by applying the *LW* formula to *Quc-q* source of gravity and electromagnetism. We begin with the latter, for which the 4-vector retarded potential at some 'field point' $x$, ($x \cdot x = \tau^2$), generated by Source $q$, is

$$Д_q^\mu(x) \equiv gQ_q \theta_{ret}(b_q, x) v_q^\mu / (v_q \cdot x - \tau), \qquad (16)$$

the retardation step function, $\theta_{ret}(b_q, x)$, being defined below via the lightcone whose vertex locates at $x$.

The symbol $\theta_{ret}(b_q, x)$ in (16) denotes an *operator* function equal to 1 *iff* the spacetime straight line of direction $v_q$, that passes at age $\tau$ through the 'source' spacetime location $x_q$, intersects the $x$ *backward* lightcone (which does not include this lightcone's vertex). Otherwise $\theta_{ret}(b_q, x)$ vanishes. (*Any* lightlike straight line *not* passing through $x$ intersects the $x$ lightcone exactly *once*.)

The Age-$\tau$ *expectation*, of the self-adjoint operator $Д^\mu(x)$ that *sums* (16) over *all qucs*, prescribes the reality-defining classical electromagnetic 4-vector field $A^\mu(x)$. The Dalembertian of $A^\mu(x)$ is the electric-charge current density—an aspect of reality that, despite 'classical' status, manifests electric-charge *discreteness*.

Discretization of *quc* energy renders almost straightforward an extension of the foregoing to gravity. In place of (16) the 9-component 'traceless' symmetric Lorentz-tensor retarded self-adjoint gravitational potential operator, that Source-$q$ generates, is defined by

$$Ƥ_q^{\mu\nu}(x) \equiv -M_q/2\tau \ \theta_{ret}(b_q, x) v_q^\mu v_q^\nu / (v_q \cdot x - \tau). \qquad (17)$$

The Age-$\tau$ expectation, of the self-adjoint operator $Ƥ^{\mu\nu}(x)$ that sums (17) over *all qucs*, prescribes the classical gravitational traceless-symmetric-tensor field $\Phi^{\mu\nu}(x)$. The Dalembertian of $-\Phi^{\mu\nu}(x)$ is the energy-momentum tensor.

There is *no* retarded SMU-*LW* 'acceleration field'. Photons are represented, within SMU reality, *not* by a zero-Dalembertian electromagnetic field but via the energy-momentum tensor--which comprises current densities of *all* energy, momentum and angular momentum.

Photon annihilations or creations—aspects of 'objective reality'--become (Popper) *physically inferrable* from *positive-energy*-momentum current density *together* with *classically-discrete* electric-charge current density—the Dalembertian of the classical but *discretely*-sourced electromagnetic vector field. Charge discreteness together with energy positivity allows experimenters to infer discrete photons from observed current densities by application of (classical) Newton-Maxwell theory to current densities of electric charge and energy-momentum.

**Self-Adjoint *Single-Quc Kinetic*-Energy Operator—a *Positive* Function of CL Casimirs**

The algebra of the 6-parameter semi-simple non-abelian *exterior*-SL(2,c) CL subgroup, comprises the conserved components of a 6-vector—a second-rank *antisymmetric* exterior-Lorentz tensor. Three algebra members associate to *quc* angular momentum, $J_q$ (a 3-vector), and three to *quc* momentum, $K_q/\tau$ (also a 3-vector). Each sextet-algebra member is represented by a self-adjoint operator on the *quc* Hilbert space. In the Dirac-coordinate ($s_q$, $y_q$, $z_q$) basis each of these momentum operators linearly and homogeneously superposes *first* (partial) derivatives. [2] As emphasized by Appendix B, each algebra member is a Dirac *momentum—not* a Dirac *coordinate*.

The two (invariant) CL-group Casimirs (commuting with all 9 of the conserved CL generators) are the 3-vector operator inner products $K_q \cdot J_q$ and $K_q \cdot K_q - J_q \cdot J_q$—homogeneous in regular-basis (partial) *second* derivatives.[2] Neither of the foregoing forms is positive, but Reference (2) displays algebraic equivalence to *another* pair of invariant self-adjoint operators, one of which has (positive-negative) *integral* eigenvalues while its companion enjoys a *continuous positive* spectrum. Denoting the former by the symbol $m_q$ and the latter by the symbol $\rho_q$, the algebraic relation is

$$K_q \cdot J_q = (\rho_q/2)(m_q/2), \quad K_q \cdot K_q - J_q \cdot J_q = (\rho_q/2)^2 - (m_q/2)^2 + 1. \qquad (18)$$

The positive *continuous*-spectrum SMU *quc*-kinetic-energy operator is $\rho_q/2\tau$, joining in the 6-element *unirrep csco* (see Appendix B) the positive *discrete*-spectrum *quc*-energy operator, $M_q/2\tau$.

**Self-Adjoint *Quc-Pair Potential*-Energy Operators**

The SMU Hamiltonian potential-energy operator is a sum over $21M_{max}(21M_{max}-1)/2$ *quc pairs* of CL-invariant electromagnetic-gravitational potential energies, $V_{qq'}(\tau) = V_{q'q}(\tau)$, whose *individual* status parallels that of the Euclidean-group-invariant 'Coulomb-gravity' potential energy in a Hamiltonian for two (slowly-moving) charged, massive particles. We postulate an SMU Hamiltonian potential-energy operator, for the *quc* pair, $qq'$, that depends on the *exterior-invariant* 'relative Dirac coordinate' $\boldsymbol{b}_{qq'} \equiv \boldsymbol{b}_q \boldsymbol{b}_{q'}^{-1}$. Notice that $\boldsymbol{b}_{qq'}^{-1} = \boldsymbol{b}_{q'q}$.

Guided by Reference (4) and the LW denominator in Formulas (16) and (17), we further postulate inverse proportionality to $e^{\beta_{qq'}} - 1$. Here the positive symbol, $\beta_{qq'} = \beta_{q'q}$, stands for (EL-invariant) shortest distance in (curved) *relative base-space* between the locations of *Quc q* and *Quc q'*. This distance equals $\cosh^{-1}[\frac{1}{2}tr(\boldsymbol{b}_{qq'}^{\dagger}\boldsymbol{b}_{qq'})]$--the *same* function of $\boldsymbol{b}_{qq'}$ as that which in Formula (13) above specified the *single-quc* coordinate $\beta_q$ in terms of $\boldsymbol{b}_q$.

Beginning with electromagnetism, as we did above when defining classical E-G fields via the *LW* formulas (16) and (17) for field operators, we postulate

$$V_{qq'}{}^{el}(\tau) = g^2 \tau^{-1} Q_q Q_{q'} (e^{\beta_{qq'}} - 1)^{-1}. \qquad (19)$$

The corresponding CL-invariant gravitational potential-energy operator is

$$V_{qq'}{}^{gr}(\tau) = -\tau^{-1} (M_q/2\tau)(M_{q'}/2\tau)(e^{\beta_{qq'}} - 1)^{-1}, \qquad (20)$$

the *complete quc*-pair potential-energy operator being the sum, $V_{qq'}(\tau) \equiv V_{qq'}{}^{el}(\tau) + V_{qq'}{}^{gr}(\tau)$. (The absence, anticipated earlier, of a separate 'nuclear-force' potential-energy Hamiltonian component will be reviewed in our concluding section.) Notice how the rightmost factor in (19) and (20) exhibits 'Newton-Coulomb' dependence on $\beta_{qq'}$ for $\beta_{qq'} \ll 1$ while 'Yukawa' exponential dependence for $\beta_{qq'} \gg 1$.

**SMU Hamiltonian and Schrödinger Equation**

As was the case for Schrödinger, our Hamiltonian sums symmetry-group-invariant self-adjoint kinetic-energy and potential-energy operators that do *not* commute. SMU dynamics proceeds through a multi-*quc* Schrödinger (first-order) differential equation where, at each post-big-bang age, a CL-invariant although age-dependent self-adjoint Hamiltonian operator (*not* a CL-algebra member) generates an infinitesimal wave-function change that prescribes the 'immediately-subsequent' universe wave function. Schrödinger's 1927 equation was similar although based on a 7-parameter extended-Euclidean group with *flat* 3-space translations (instead of the 9-parameter CL group with *curved* 3-space translations).

The invariant age-dependent self-adjoint Hamiltonian operator is

$$H(\tau) = \Sigma_q\, \rho_q/2\tau + \Sigma_{q \neq q'} V_{qq'}(\tau), \qquad (21)$$

while the evolution equation for the universe ray is

$$i\partial \Psi(\tau)/\partial \tau = H(\tau)\, \Psi(\tau) . \qquad (22)$$

An *initial* ray of *uncorrelated* (bachelor) *quc*s, at $\tau = \tau_0$, is proposed in Appendix A. Notice that our Schrödinger equation, in *absence* of gravitational potential energy, is conformally ('scale') invariant—dependent only on age *ratios* and thereby paralleling a QFT feature important to renormalization. Related is our conjecture that Maxwell's equations are satisfied by SMU classical electromagnetic fields.

**Conclusion**

A 9-parameter 'centered-Lorentz' (CL) Lie symmetry group collaborates with a Schrödinger equation that prescribes Schrödinger-Milne quantum-universe electro-gravitational evolution with increasing universe *age*. SMU resides *inside* a *forward* lightcone, the age of any location its 'Minkowski distance' from lightcone vertex. Age is a CL-invariant *nongeometrical* perpetually-increasing parameter approximately equal at present to the reciprocal of Hubble's 'constant'.

At each age greater than or equal to a starting (big-bang) age (1, in units where $G = \hbar = c = 1$; in *seconds*, big bang age is $\sim 10^{-43}$), the Dirac-coordinate-basis universe ray is a sum of (tensor) products of single-*quc* wave functions. A '*quc*' is an SMU constituent. The *argument* of a *quc*'s wave function in the latter's 'Dirac-coordinate' basis specifies the *quc*'s location within a CL-dictated 6-dimensional manifold.

The number of *quc*s is ginormous but finite and unchanging. Each *quc* represents CL [Formula (8)] through displacements of its 6 Dirac coordinates *and* of its wave-function phase. The total *number* of *quc*s and the unit of electric-charge remain to be specified. An estimate of the latter is provided by the (incomplete) 'Standard Model of particle-physics'.

Appendix A proposes an initial ray of 'bachelor' *quc*s—*devoid*, at the beginning, of mutual correlations. We suppose the first 'marriages' of electrically-charged *quc*s to have emerged at GUT-scale ages ($\sim 10^{-39}$ sec) with creation of 2-*quc* 'double-helix cosmological photons'. Later, at micro-scale universe ages ($\sim 10^{-24}$) sec, there emerged 'massive elementary particles—electrically-neutral 2-*quc* neutrinos, Higgs bosons and $Z_0$'s, together with charged 3-*quc* quarks, leptons and W bosons. (Charged-*quc* composition of elementary particles, broached in Appendices C and E, will be addressed in separate papers.) At macro-scale universe ages ($\sim 10^{-5}$ sec), we believe stellar construction began. All such conjectures are in principle verifiable by computation.

Present-age ($\sim 10^{17}$ sec) 'dark matter' comprises galactic-scale colonies of electrically-neutral 'bachelor' *quc*s that gravitationally attach to entire stellar galaxies. A (still larger) universe component, remaining today galaxy-*unassociated*, comprises *quc* bachelors that so far have maintained their independence.

*Quc* chirality (conjugate to *one* of six *quc* Dirac coordinates) we have ('Occam') limited to the values 0,±1—a 'Dirac-tripling' that accompanies limitation on any *quc*'s electric-charge integer to the 7 values 0, ±1, ±2, ±3. In early-universe dynamics we believe the combination of chirality and electric charge at particle-physics micro scale to have (dynamically) distinguished baryon-number-carrying *quarks* and associated 'strong interactions' from elementary bosons and leptons with zero baryon number (Appendix, Table I).

We expect the SMU Schrödinger equation (22) to reveal 'nuclear forces', along with other particle-physics, as an *approximate* notion—useful at *micro scale* but *not* at *all* SMU scales and not a *foundational* feature of a *quantum* universe--*all* of whose 'forces' our Hamiltonian proposes to be electro-gravitational. The 'short range' of nuclear forces manifests electric-charge *screening*—important whenever the *number* of *quc*s in some *charged-quc* set exceeds the *sum* of this set's charge integers.

*All* physics measurements 'Popper-rely' for interpretation of observed objective reality, on classical-physics electro-gravitational theory. The SMU ray specifies, through expectations of self-adjoint electromagnetic and gravitational field operators, a physics-enabling 'fixed and settled reality' that includes locally-*unobservable* energy (Bohm hidden reality) *together with* macro-scale '*observable* objectivity'.

**Acknowledgements**.

Decades of discussions with Henry Stapp have been invaluable to this paper. Also contributing to the ideas here have been Eyvind Wichmann, David Finkelstein, Jerry Finkelstein, Dave Jackson, Stanley Mandelstam, Ralph Pred, Bruno Zumino, Ramamurti Shankar, Don Lichtenberg, Ling-Lie Chau, Ivan Muzinich, Korkut Bardacki, Bob Cahn, Lawrence Hall, and Nicolai Reshetikhin.

Essential has been support and encouragement, especially during the final years of my life, from my 5 children, Pauline, Frank, Pierre, Beverly and Berkeley. Without their participation this creation would never have occurred.

**Appendix A. Initial-Universe (Planck-Scale) Wave Function**

The *initial* ($\tau = \tau_0 = 1$) SMU wave function we propose to have been a *single* product of $21 M_{max}$ *single-quc* wave functions that represent *uncorrelated* 'initially-bachelor' *quc*s. No (GUT-scale or *micro*-scale) marriages between 2 or 3 *quc*s—electro-gravitationally-stabilized elementary particles--were *a priori*. All particles and dark matter--*all* objective reality—we presume to have resulted from Schrödinger-equation Hamiltonian-generated *evolution*. *Present*-universe Hubble scale—the spacetime scale set by *our* universe's age--is larger than the scale of $\tau_0$ by a factor of order $10^{60}$.

Von Neumann ideas have led us to propose, as initial ray in the regular CL-representation ('Dirac-coordinate') basis, the following product of $21M_{max}$ 'gaussian' factors, with *no* dependence on the *arguments* of the complex coordinates $y_q$ and $z_q$—dependence *only* on their *magnitudes*,

$$\Psi(\tau_0) = \Pi_q \, exp \, (-iN_q \, Im \, s_q) \, exp \, (-iM_q \, Re \, s_q) \, |y_q z_q|^{-1} exp \, (- \tfrac{1}{2} ln^2|y_q|- \tfrac{1}{2} ln^2|z_q|). \qquad (A.1)$$

With the starting ray (A.1), which recognizes electric charge by the subscript $q$ on any GN-Dirac coordinate being *defined* as synonymous with the integer trio $Q_q$, $N_q$, $M_q$, the 3-vector momentum operator of each *quc* has vanishing *expectation*. This ray, further, is an *eigenvector* of total 3-vector *angular* momentum with *zero* eigenvalues for all components thereof—thereby satisfying classical Mach-Milne principles perpetuated by (main-text) Eq. (22). *Total* electric charge and chirality have zero eigenvalues in the initial ray (A.1) as well as in all subsequent rays. Total starting energy has the value, $21(M_{max}/2)^2 \tau_0^{-1}$. Subsequent total SMU energies replace $\tau_0$ (=1) in the foregoing formula by $\tau$.

**Appendix B. CL Unirrep-Csco as 'Dirac Momentum' Basis**

GN's unitary $SL(2,c)_R$ Hilbert-space representation via normed complex differentiable functions, $\Psi(a)$, of (single-*quc*) location within the 6-dimensional group manifold, was called by these authors the 'regular' Lorentz-group representation. [2] Our main text has characterized the corresponding complete set of 6 commuting self-adjoint operators ('csco') as a 'Dirac-coordinate' Hilbert-space basis. *Not* GN-emphasized is representability, by this *same* csco, of the 12-parameter group $SL(2,c)_L \times SL(2,c)_R$. An 8-parameter subgroup of the latter—keeping only the *diagonal left*-multiplying 2×2 complex-unimodular matrices--is the electric-charge-ignoring subgroup of the 9-parameter SMU-foundational CL group. [4] [The CL center *augments* $SL(2,c,D)_L$ with $U(1)$.]

Exposed in detail by GN, *beyond* their regular $SL(2,c)_R$ representation, was the latter group's unitary *irreducible* representation by a 'unirrep-csco' that *also* [although not noted in Reference (2)] represents 12-parameter $SL(2,c)_L \times SL(2,c)_R$. GN's 'Lorentz unirrep' becomes a 'Dirac-momentum-basis' when 2 of its 6 csco members are 'Fourier-replaced' by CN-center generators. (The 4-member remainder of GN's 6-member *unirrep*-csco commutes with the foregoing *pair* of Dirac-momentum csco members.)

*Both* the GN unirrep-csco *and* its 'Dirac-momentum counterpart' include the two Lorentz-group Casimirs appearing in the main-text paragraph preceding Formula (18). These Casimirs commute with all 12 generators of the 'left-right Lorentz group' and with all 9 generators of the CL group. The pair of self-adjoint operators appearing in the Dirac-momentum csco (but *not* in GN's *unirrep* csco) are the main-text energy and chirality integer-eigenvalued operators. CL unirrep then follows from 6-dimensional unirrep-csco augmentation by electric charge.

The complete set of 7 commuting self adjoint 'Dirac-momentum' *single-quc* operators comprises $Q_q$, $N_q$, $M_q$, $m_q$, $\rho_q$, and $z_{1q}$, where $Q_q$ and $N_q$ take (the by now familiar) possible values $Q_q = 0, \pm 1, \pm 2, \pm 3$ and $N_q = 0, \pm 1$. The integer $m_q$ takes *all* positive-negative-integer values while $\rho_q$ has a continuous spectrum spanning the *positive* real line and $M_q$ takes (main-text, positive-integer) values $1, 2 \ldots M_{max}$. Evenness (oddness) of $m_q$ is accompanied by evenness (oddness) of $N_q$. Earlier we have described $\rho_q/2\tau$ as the 'local-frame kinetic energy of *Quc q*'; in the present context, 'local-frame magnitude of *Quc-q* momentum' is a more appropriate appellation. (Remember that '*quc* mass' is devoid of meaning!)

The continuous spectra of both $Re\, z_{1q}$ and $Im\, z_{1q}$ span (full) real lines; we choose to describe the 'meaning' of $z_{1q}$ as 'local-frame-direction of *Quc-q* 3-vector momentum'. Reminder of trickiness in terminology choice is the Haar measure for the 3-dimensional (noncompact) $\rho_q$, $z_{1q}$ subspace: $(m_q^2 + \rho_q^2)\, d\rho_q\, dz_{1q}$. [2] [In terms of directional (real) polar angles, $\pi \geq \theta_{1q} \geq 0$ and $2\pi > \varphi_{1q} \geq 0$, the complex $z_{1q}$ may be written $i\, tan(\theta_{1q}/2)\, exp(i\varphi_{1q})$. The absolute value of $z_{1q}$ is then $tan(\theta_{1q}/2)$.]

We perceive the set of 4 commuting self-adjoint operators, $m_q$, $\rho_q$, and $z_{1q}$, as 'Dirac-momenta' even though *not* 'canonically conjugate' to the $y_q$, $z_q$ quartet of Dirac coordinates—the complex commuting operators that, despite lack of 'ordinary-language' names, have appeared prominently in this paper's main text. We think of $m_q$ as '*quc* helicity'—the 'component of *quc* angular momentum in the direction of its momentum'. The author is comfortable in calling the GN-Dirac coordinate $z_q$ '*quc* velocity direction' (related to velocity polar angles in the manner above used for 'momentum direction') but has yet to achieve comfort with *any* (physics-familiar) name for the GN-Dirac coordinate $y_q$.

**Appendix C. Photons of *Differing* Diameters Although *Same* Momentum and Helicity**

Main-text noted has been SMU's (2-*quc*) *4-dimensional* (3-momentum plus helicity) 'Dirac-momentum' basis for the 'external' properties of a *single* photon. But $\gamma_c$ *also* has an 'internal' Hilbert vector—a function of the 2-*quc relative* coordinate, whose *spatial* extension transverse to momentum direction might be called 'photon diameter'.

The *internal* $\gamma_c$ Hilbert vector is a complex normalizable function of location within a Dirac-*relative*-coordinate manifold. Among elementary particles, photons are 'special' by important dependence of their internal *quc* dynamics on *gravitational* attraction between *quc*s, as well as on electromagnetic inter-*quc* attraction or repulsion.

The foregoing we have main-text summarized by attaching to the photon the acronym, 'double helix'. A physics-unappreciated photon attribute is GUT-scale double-helix diameter.

With (see Table I below) $Q_q = \pm 3$ and $Q_{q'} = -Q_q$, (local-frame) $qq'$-composed photon energy at age $\tau$ is $E_\gamma = (2\tau)^{-1}(M_q + M_{q'})$. This 'external' energy remains unchanged if $M_q \rightarrow M_q + \sigma$, $M_{q'} \rightarrow M_{q'} - \sigma$, with $\sigma$ an integer whose absolute value is smaller than either $M_q$ or $M_{q'}$. Gravitational potential energy--$V_{qq'}^{gr}$—by Formula (20), however, *is* changed, with an associated change in double-helix diameter. (Plausibly the *smallest* diameter associates to $M_q = M_{q'}$.)

The total number of *different* double-helix *quc* pairs, with *same* momentum and helicity (and same chirality) but *differing* helix diameters, is $\tau E_\gamma$. Even for 'soft' photons in the present universe (those of wavelength ~ km or greater), the number of *different* SMU ($\gamma_c$) photons sharing the *same* momentum and helicity is of order $10^{22}$. (A huge number have 'almost the same' diameter.)

It follows that, despite SMU *finiteness* of photon total number, high accuracy may attach to physics *coherent-state* QFT representation of classical electromagnetic radiation—a $\tau \rightarrow \infty$ approximation that recognizes *indefinitely* many FHPP-*identical* photons. Present-universe FHPP accuracy of Bose-Einstein *identical-photon* statistics is understandable even though any $\gamma_c$ is *different* from any *other*. *Identity* of all 'photons with common momentum and helicity' is one of many physics approximations that accompany 3-space flattening.

Already at spatial *micro*-scales (far above GUT scale although far below macro scale) physics notions such as Bose-Einstein and Fermi-Dirac statistics become accurate. Lack of meaning for '*quc* statistics' accompanies *higher* dimensionality of '*quc* space' compared to that of 'particle space'.

**Appendix D. Dark Matter as Non-Particulate (Bohm) 'Hidden Reality'**

Via a self-adjoint Hamiltonian with kinetic- and potential-energy components, the universe's evolving 'ray' specifies 'evolving reality' through expectations of self-adjoint operators that represent current densities of energy, momentum, angular momentum and electric charge. A 'particle' is a *micro*-scale *clump* of energy-momentum with some *integral* electric charge, some baryon number, some integral or half-integral angular momentum in units of $\hbar$ and some approximately-determined mass. *All qucs* in particle clumps are electrically charged.

Any 'observer' is a *macro*-scale clump of particles with *approximately-zero* total charge (charge screening). Distinction between 'particle' and 'observer' resides not in the SMU Hamiltonian but in ray aspects that emerge as distinct scales develop with universe expansion. 'Dark matter' resides in *non-particulate galactic-scale* clumps of chargeless but energy-carrying *quc*s. Neither particle nor observer 'contains' dark matter.

Schrödinger's equation determines reality evolution without requiring either that all energy density be 'particulate' or that all reality be *micro-macro*-scale. *Galactic*-scale 'dark matter' is the *non-particulate* electrically-neutral source of gravitational potential energy which helps *determine*, via the SMU Schrödinger equation, the age at which a 'radioactive' particle decays. Dark matter constitutes reality that Bohm characterized as 'hidden'. *All* SMU history is 'deterministic'—with 'observations' merely one among many 'onflow' (Ralph Pred's term) aspects.

**Appendix E. 3 Elementary-Fermion Generations**

The famously-mysterious 3 generations of QFT elementary fermions associate to the three possible absolute values for nonvanishing *quc* electric charge. (*All* QFT elementary particles, whether charged or neutral, are composed *exclusively* of *charged qucs*.) Generation mass magnitudes we believe associate *inversely* to the *quc*-charge integers 1, 2, 3. Elementary-fermion mass ratios are, to an accurate approximation, *electrodynamically* determined. Approximate mass-ratio gravity-ignoring computation should be possible.

Each of the three *qucs* (approximately) building *any* charged elementary fermion carries a charge integer, ±1, ±2, ±3. *Two* of the three *individually* carry *zero* chirality while *opposite* charges--the *net* chirality *and* charge of this "core' pair *vanishing*. The remaining 'valence'-*quc* carries the fermion's chirality, charge and spin as well as baryon number. For a charged lepton the charge, is $-3g$ while the (fluctuating) chirality is $\pm\hbar/2$. For a quark the valence *quc* carries either the charge $2g$ or the charge $-g$, together with baryon number ⅓ and $\pm\hbar/2$ (fluctuating) chirality.

Three generations of charged elementary fermions associate to 3 possibilities for the electrically-neutral 'core'-pair of *individually-zero-chirality* charged *qucs*. We expect the lowest-mass generation to be that with the $Q = \pm 3$ pair because here the *negative* electromagnetic potential energy is greatest. The highest-mass generation we expect to be that with the $Q = \pm 1$ *quc* pair.

Three 'types' of neutrino, each a $Q = \pm 1$, $\pm 2$ or $\pm 3$, *quc* pair, differ from the foregoing 3 charged 'generations' by *one* member of any neutrino *quc*-pair having $N = \pm 1$, with a sign that in our galaxy *agrees* with that of this *quc*'s electric charge; the *other quc* has zero chirality.

**Table I**.

A pedagogically-helpful definition is possible, through *quc* electric charge, of SMU 'dark side' and 'bright side', separated by a 'baryonic middle'. Note that *qucs* composing photons, electrons, positrons and first-generation neutrinos (Dirac's concern) are exclusively 'bright'.

|       | Dark | Baryonic |      | Bright |
|-------|------|----------|------|--------|
| $Q_q$ | 0    | ±1       | ±2   | ±3     |
| $B_q$ | 0    | ∓⅓       | ±⅓   | 0      |